\documentclass{osa-article}

\journal{osac}

\articletype{Research Article}

\usepackage{siunitx}

\begin{document}

\title{Self-stabilization mechanism in ultra-stable Fourier domain mode-locked (FDML) lasers}

\author{Mark Schmidt,\authormark{1,*} Tom Pfeiffer,\authormark{2} Christin Grill,\authormark{2} Robert Huber,\authormark{2} and Christian Jirauschek\authormark{1}}

\address{\authormark{1}TUM Department of Electrical and Computer Engineering, Technical University of Munich, Arcisstra\ss e 21, 80333 Munich, Germany\\
\authormark{2}Institute of Biomedical Optics, 
University of L\"ubeck, Peter-Monnik-Weg 4, 23562 L\"ubeck, Germany}

\email{\authormark{*}mark.schmidt@tum.de} 



\begin{abstract}
Understanding the dynamics of Fourier domain mode-locked (FDML) lasers is crucial for determining physical coherence limits, and for finding new superior methods for experimental realization. In addition, the rich interplay of linear and nonlinear effects in a laser ring system is of great theoretical interest. Here we investigate the dynamics of a highly dispersion-compensated setup, where over a bandwidth of more than \SI{100}{\nano\meter}, a highly coherent output with nearly shot-noise-limited intensity fluctuations was experimentally demonstrated. This output is called the sweet-spot.
We show by numerical simulation that a finite amount of residual dispersion in the fiber delay cavity of FDML lasers can be compensated by the group delay dispersion in the swept bandpass filter, such that the intensity trace exhibits no dips or high-frequency distortions, which are the main source of noise in the laser. In the same way, a small detuning from the ideal sweep filter frequency can be tolerated. Furthermore, we find that the filter's group delay dispersion improves the coherence properties of the laser, and acts as a self-stabilizing element in the cavity. Our theoretical model is validated against experimental data, showing that all relevant physical effects for the sweet-spot operating regime are included.      
\end{abstract}

\section{Introduction}
Fourier domain mode-locking is a laser operating mode in which a tunable optical bandpass filter is swept synchronously to the roundtrip time of the optical field in the laser cavity. The main components are a rapidly tunable Fabry-P\'{e}rot (FP) filter with moving mirrors and a fiber delay line in the order of several hundreds of meters to kilometers. This laser principle has no upper limit to the tuning speed under ideal resonance conditions~\cite{jirauschek2015wavelength} since the laser remains in single-mode operation. The change in the wavelength in this case is obtained by the Doppler shift, caused by the moving mirror of the FP optical bandpass filter~\cite{jirauschek2015wavelength,SiegmanLasers}. Thus, the laser tuning does not suffer from mode-hopping, which is the major limiting factor in the tuning speed of swept laser systems. FDML
lasers have revolutionized optical imaging and sensing techniques, especially optical coherence tomography (OCT)~\cite{huang1991optical,wieser20105MHzOCT,biedermann2008real,adler2011extended,reznicek2014megahertz,klein2013joint,biedermann2009recent}, and applications in FDML-based Bragg grating sensor systems \cite{jung2008characterization,lee2011remote,lee2010dynamic} have been realized. Furthermore, real time spectroscopy~\cite{kranendonk2007high,kranendonk2005modeless,kranendonk2007wavelength}, high-speed stimulated Raman scattering spectroscopy~\cite{karpf2015time} and picosecond pulse generation \cite{eigenwillig2013picosecond} have been demonstrated. The combination of high output powers in the range of~\SI{100}{\milli\watt}~\cite{mao2009HighOutputPower}, sweep frequencies of up to \SI{5}{\mega\hertz}~\cite{wieser20105MHzOCT}, typical sweep bandwidths of more than \SI{100}{\nano\meter} and instantaneous linewidths below \SI{0.1}{\nano\meter}~\cite{adler2011extended} make the FDML laser one of the favored systems for tunable lasers. Other important swept-laser sources are vertical-cavity surface emitting lasers with a microelectromechanical system (MEMS-VCSEL) \cite{qiao2017VCSELReview,potsaid2012mems}, and Vernier-tuned distributed Bragg reflector (VT-DBR) lasers~\cite{bonesi2014akinetic}.\\
In their normal operation mode, FDML lasers suffer from high intensity noise, with fluctuations of almost \SI{100}{\percent}~\cite{pfeiffer2017analysis}. The intensity fluctuations constitute high-frequency noise compared to the sweep speed of the laser wavelength, and significantly influence the phase stability and coherence length of the laser. A main limiting factor is considered to be chromatic dispersion in the fiber cavity, which delays or accelerates the instantaneous frequency (IF) of the optical field with respect to the sweep period of the FP filter during each pass through the fiber delay line, and accumulates over consecutive roundtrips unless there is a reverse shift.\\ 
A new operating regime discovered by Kraetschmer and Sanders \cite{kraetschmer2009ultrastable}, called the sweet-spot, showed shot-noise-limited intensity fluctuations and a significantly narrowed instantaneous linewidth in a \SI{3.5}{\nano\meter} band. This was achieved via a dispersion-compensated setup, with near-perfect synchronization of the sweep frequency to the optical roundtrip time, in the order of a few \SI{}{\milli\hertz}. This ultra-stable operating mode was recently extended to \SI{117}{\nano\meter}~\cite{pfeiffer2018ultra} and paves the way towards reaching the coherence limit in FDML lasers. In particular, such low-noise light sources are essential for OCT systems in order to achieve shot-noise-limited detection sensitivity for high-quality images, which has already been observed in former time domain OCT systems~\cite{hartl2001ultrahigh,nishizawa2004real}. It is even more critical in recent swept-source OCT systems, since here not only the system sensitivity but also the maximum ranging depth~\cite{pfeiffer2018ultra} and the dynamic range~\cite{huber2006buffered} can be negatively affected by laser noise.\\
An understanding of the rich FDML laser dynamics is therefore of fundamental importance for reaching the coherence limit. Here we present a self-stabilization mechanism in a highly dispersion-compensated setup, where a finite remaining dispersion in the fiber delay cavity can be compensated due to the dynamics induced by the frequency-dependent group delay of the swept FP filter.  
This self-stabilizing effect leads to a tolerance over non-synchronized resonance conditions in the ring laser setup, which ultimately makes sweet-spot operation possible at all. We believe that understanding and controlling this self-stabilization effect can improve FDML frequency combs, which were demonstrated in\cite{tsai2009frequency,wan2017time,lippok2019extended}, and might enable short pulse generation in the low- or sub-\SI{}{\pico\second} range.\\
The nonlinear system dynamics of FDML lasers has been studied theoretically within different approaches, whereby the problem was modeled as a time-delay dynamical system~\cite{pimenov2017dispersive,slepneva2013dynamics,slepneva2019convective} or with the real-valued Ginzburg-Landau equation with a zero group delay (GD) Gaussian filter~\cite{li2017eckhaus}. Within these approaches, several stability boundaries were derived by a linear stability analysis of stationary solutions, and different operating regimes such as the Turing~\cite{slepneva2013dynamics}, the Eckhaus~\cite{li2017eckhaus} or a complex modulation instability~\cite{pimenov2017dispersive,slepneva2013dynamics,slepneva2019convective} were discussed. As far as we are aware, none of these investigations refer to a stabilizing mechanism which is due to the dispersive group delay in the FP filter, enabling us to identify a physical mechanism behind the stability boundaries. Furthermore, in~\cite{li2017eckhaus} it is found that the dispersion causes an accumulative frequency shift which will always drive the laser out of the stability region. Note that in this model, a non-causal Gaussian filter was used which has no frequency-dependent group delay.\\
In the following, we discuss the laser setup for ultra-stable operation and the underlying physical model. Furthermore, we present the self-stabilization mechanism and validate our model against experimental data. 
 
\section{Laser setup for ultra-stable operation} \label{sec_ExpSetup}
\begin{figure}[t]
\centering
\includegraphics[width=30pc]{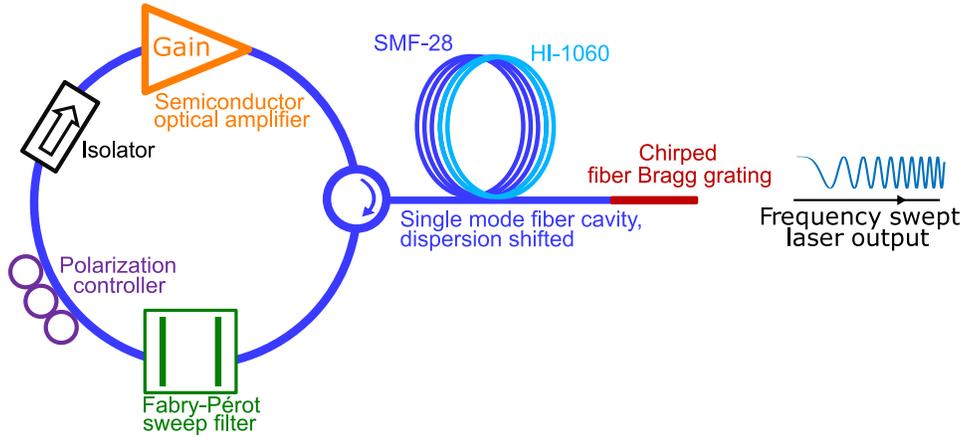}
\caption{Schematic illustration of the FDML laser setup for ultra-stable operation.}
\label{fig_FDML_Setup}
\end{figure}

The FDML laser system investigated here~\cite{pfeiffer2018ultra} is shown in Fig.~\ref{fig_FDML_Setup}. It consists of a FP tunable bandpass filter and a single-polarization semiconductor optical amplifier (SOA) gain medium which causes the optical field to have a dominant, linearly polarized component. The polarization controller is therefore used in experiment to manually align the polarization state of the optical field with the SOA gain axis. The fiber delay line is dispersion-shifted by mixing two fibers, a SMF-28 and a HI-1060 fiber, in order to move the zero-dispersion wavelength out of the sweep range. In this case, a chirped fiber Bragg grating (cFBG) can be used in order to compensate for chromatic dispersion which was manually fine-tuned by a temperature gradient. The cFBG serves at the same time as the outcoupling element. The isolator ensures unidirectional lasing and the whole laser is temperature controlled, which is critical for reaching the sweet-spot operation mode in addition to a high sweep frequency stability. The sweep filter frequency was automatically regulated in order to reach sweet-spot operation and to adjust the laser to environmental changes such as temperature fluctuations or drifts in the local oscillator generating the sweep frequency. Further details can be found in~\cite{pfeiffer2018ultra}.
\section{Modeling of ultra-stable operation}
The FDML laser dynamics is governed by the field envelope transformations of each of the various components in the laser system. Therefore, we follow a lumped-element approach based on the models in~\cite{jirauschek2009theoretical} and \cite{jirauschek2017efficient} but with a physical model for the SOA dynamics instead of a phenomenological one. As in our previous investigations of FDML lasers, we use the swept-filter reference frame in which the IF of the optical field is modulated with the center frequency of the sweep filter~\cite{jirauschek2009theoretical}. Within this approach, the optical field is comoving with the transmission peak of the FP bandpass filter and ``sees'' a static bandpass filter instead of a swept one. The variations from the peak transmission are thereby contained in the optical field. This approach reduces the simulation bandwidth drastically by three orders of magnitude~\cite{jirauschek2009theoretical} because the simulation bandwidth is limited by the bandwidth of the sweep filter and not by the whole optical sweep bandwidth. For comparison, a monochromatic field would correspond to a swept field in the reference frame.  
\\
The fiber delay cavity including the cFBG modulates the phase of the complex envelope according to
\begin{align} \label{eqn_fiber}
u_\mathrm{out,fiber}\left(\tau\right) = \sqrt{\kappa_\mathrm{f} }\, u_\mathrm{in,fiber}\left(\tau\right) \exp&\left\{ \mathrm{i}L_\mathrm{f} \left[\frac{1}{2}\beta_2\omega_s^2\left(\tau\right) +  \frac{1}{6}\beta_3\omega_s^3\left(\tau\right) +  \frac{1}{24}\beta_4\omega_s^4\left(\tau\right)\right] +\mathrm{i}\varphi_\mathrm{cFBG}(\tau)
\right.\nonumber \\ &  \left. {} + \mathrm{i} \gamma \kappa_\mathrm{f} \bigl| u_\mathrm{in,fiber}\left(\tau\right)\bigr| ^2\frac{1}{2}L_\mathrm{f}\left(1 + R \right) \right\} 
\end{align}
where $L_\mathrm{f}$ is the propagation length in the single-mode fiber, $\beta_2$ to $\beta_4$ are the second- to fourth-order dispersion coefficients, $\gamma$ is the nonlinearity coefficient of the fiber and $\kappa_\mathrm{f}$ accounts for the power loss in the fiber spool, including splicing, circulator as well as fiber losses. For the nonlinear term in the exponential of Eq.~(\ref{eqn_fiber}), $\kappa_\mathrm{f}$ is for simplicity fully considered because the exact distribution of losses has not been experimentally characterized. $R$ is the reflectivity of the cFBG. The fiber length $L_\mathrm{f}$ takes forward and backward propagation after reflection at the cFBG into account. The center frequency of the swept filter is $\omega_s\left(\tau\right) + \omega_c $ with the center frequency of the sweep $\omega_c$ where the relative sweep frequency $\omega_s\left(\tau\right)$ is chosen here to be $0.5 D_\omega\cos \left(\omega_0 \tau \right)$ with the sweep bandwidth $D_\omega$ and the sweep filter tuning frequency $\omega_0 $ but can in principle be an arbitrary periodic function. We abbreviate the complex field envelope at the spatial fiber input $u_\mathrm{in,fiber}\left(z=z_\mathrm{in,fiber},\tau\right) $, or output $u_\mathrm{out,fiber}\left(z=z_\mathrm{out,fiber},\tau\right) $ respectively, with $u_\mathrm{in,fiber}\left(\tau\right)$ or $u_\mathrm{out,fiber}\left(\tau\right)$ where we always refer to the spatial input or output position inside the cavity, and drop the $z$ dependence in the following. The cFBG is modeled as an ideal dispersion-compensating element by partially reverting the phase accumulation due to the fiber dispersion by $\varphi_\mathrm{cFBG}(\tau) = -L_\mathrm{f} \left[\left( 1/2 \right) \hat{\beta}_2\omega_s^2\left(\tau\right) +  \left( 1/6 \right) \hat{\beta}_3\omega_s^3\left(\tau\right) +  \left( 1/24 \right) \hat{\beta}_4\omega_s^4\left(\tau\right)\right] $. In this way, the dispersion relation of the cFBG needs not be implemented explicitly, but can be included by simply scaling the dispersion coefficients of the optical fiber by $\beta_i-\hat{\beta}_i = \beta_i/c_\mathrm{scale} $ with $ i \in \left\{2,3,4\right\}$ and a constant $c_\mathrm{scale}$ in order to simulate a residual dispersion in the case of non-ideal compensation.      
\\
The SOA gain medium modifies the complex field envelope according to
\begin{equation} \label{eqn_SOA_FieldUpdate}
u_\mathrm{out,SOA}\left(\tau\right) = u_\mathrm{in,SOA}\left(\tau\right) \exp\left\{ 0.5 h\left(\tau\right) \left[ 1 - \mathrm{i} \alpha  \right]  \right\}
\end{equation}
where  $ h\left(\tau\right) = \int_{0}^{L}g\left(z,\tau \right) \mathrm{d}z $ is the integrated power-gain coefficient $g\left(z,\tau\right)$ over the length $L$ of the gain medium, and $\alpha$ is the linewidth enhancement factor responsible for self-phase modulation (SPM) in the saturated amplifier. The integrated gain has the dynamics \cite{agrawal1989self}
\begin{equation} \label{eqn_SOA_ODE}
\frac{\partial h\left(\tau\right) }{\partial \tau} = \frac{h_0\left[ w_\mathrm{s}\left(\tau\right)+ \omega_c  \right] - h\left(\tau\right)}{\tau_c} - \frac{\bigl| u_\mathrm{in,SOA}\left(\tau\right) + u_\mathrm{ASE}\left(\tau\right)\bigr| ^2 }{P_\mathrm{sat}\left[ w_\mathrm{s}\left(\tau\right)+ \omega_c \right] \tau_c } \left[ \mathrm{e}^{h\left(\tau\right)} -1 \right]
\end{equation}
with the saturation power $P_\mathrm{sat}$ and the carrier lifetime $\tau_c$. The amplified spontaneous emission (ASE) noise $u_\mathrm{ASE}\left(\tau\right)$ is modeled as additive white Gaussian noise with zero mean and effective standard deviation $ \sigma = \left( 0.5P_\mathrm{n}\Delta_\mathrm{sim}/G_\mathrm{eff}\right)^{1/2} $. Here $P_\mathrm{n}$ is the ASE noise power within the gain bandwidth, $G_\mathrm{eff} = \int_{0}^{\infty}\exp\left[h_0\left(f\right) \right]\mathrm{d}f $ and $\Delta_\mathrm{sim}$ is the simulation bandwidth. The factor $0.5$ accounts for the real and imaginary part of the complex field envelope. Note that the optical field envelope $u\left(\tau \right)$ is normalized to the optical power and has the unit $\sqrt{\SI{}{\watt}}$. The unsaturated static gain $\exp\left[h_0(\omega)\right]$ and the saturation power $P_\mathrm{sat}\left(\omega\right)$ are frequency-dependent and are modulated in the swept-filter reference frame with the sweep frequency of the filter, which is a good approximation for small deviations of the IF of the optical field from the filter's sweep frequency. The unsaturated gain curve as well as the saturation power were measured (Thorlabs BOA1132S) and are interpolated in the simulation. The frequency dependency of $\alpha$ and $\tau_c$ is neglected here, in order to reduce the complexity of the underlying physical model.
\\
As mentioned above, the sweep filter in the swept-filter reference frame is  represented by a linear static filter which is treated in the frequency domain~\cite{jirauschek2015wavelength,jirauschek2009theoretical}. With a Lorentzian lineshape of width $\Delta_\omega$ we obtain for the field transformation of the filter per roundtrip
\begin{equation} \label{eqn_FP_update}
U_\mathrm{out,FP}\left(\omega \right) = U_\mathrm{in,FP}\left(\omega\right)H\left(\omega \right)
\end{equation}
where
\begin{equation} \label{eqn_H_w}
H\left(\omega \right) = \frac{\sqrt{T_\mathrm{max}}}{1 - \mathrm{i}2\omega/\Delta_\omega }
\end{equation}
and $T_\mathrm{max}$ is the maximum power-transmission factor.
\\
Polarization effects in the fiber spool are neglected in our model, since the amplifier is a single-polarization device, causing the degenerate orthogonal fiber mode to have a minimal influence on the laser dynamics. However, it should be noted that in principle, a passive stabilization effect due to polarization dynamics could be beneficial for the ultra-stable operating regime~\cite{jirauschek2017efficient}. Furthermore, we do not take into account spectral hole burning or carrier heating in the SOA, since we are only interested in the dominant effects in the gain medium, for the sake of simplicity. Nevertheless, such effects could easily be added in a quasi-static approach. \\

\section{Stability limit for non-synchronized resonance conditions} 
The ideal laser resonance condition for an FDML laser requires the roundtrip time of the optical field to coincide with the sweep period of the tunable bandpass filter for each instantaneous wavelength of the optical field~\cite{jirauschek2015wavelength}. Due to the wavelength dependence of the laser components such as fiber dispersion, the frequency-dependent group delay of the optical bandpass filter, or nonlinear frequency shifts arising from SPM in SOA and fiber, this condition cannot be achieved for a constant sweep period unless there exists a frequency-dependent compensation mechanism.\\
When the laser is operating outside its stability range, a main source of noise in the intensity trace of the laser output are dips (also refered to as holes~\cite{pfeiffer2018ultra,pfeiffer2017analysis}) whose occurrence appears to be irregular. Such hole formations have experimentally been observed in FDML lasers in~\cite{slepneva2019convective} and are associated with Nozaki-Bekki holes which are spatially localized solutions of the one-dimensional complex Ginzburg-Landau equation~\cite{bekki1985formations}. The stability of these holes is sensitive to perturbations and the whole system undergoes bifurcation points, separating stability regions which have complex dependencies. More details are discussed for example in~\cite{chate1992stability,chate1994spatiotemporal}. A mathematical proof of whether such a solution is also valid within the FDML laser equations is yet to be provided. \\
In this work, we find - by numerical simulation - that the laser can operate without holes in the intensity trace within a certain tolerance of either a residual dispersion or a small detuning from the optimal sweep filter frequency. We consider the optimal sweep filter frequency as the inverse of the roundtrip time which is $L_\mathrm{f}/v_\mathrm{g}$ where $L_\mathrm{f}$ is the fiber propagation length and $v_\mathrm{g}$ is the group velocity in the fiber, i.e. for an ideal delay fiber, and no delay within the other components. In the absence of other delays, this would result in an optical field with an IF which is zero in the swept-filter reference frame and thus matches the center of the filter transmission function for all times. 

\subsection{Dispersion compensation} \label{sec_DispComp}
\begin{figure}[t]
\centering
\includegraphics[width=\textwidth]{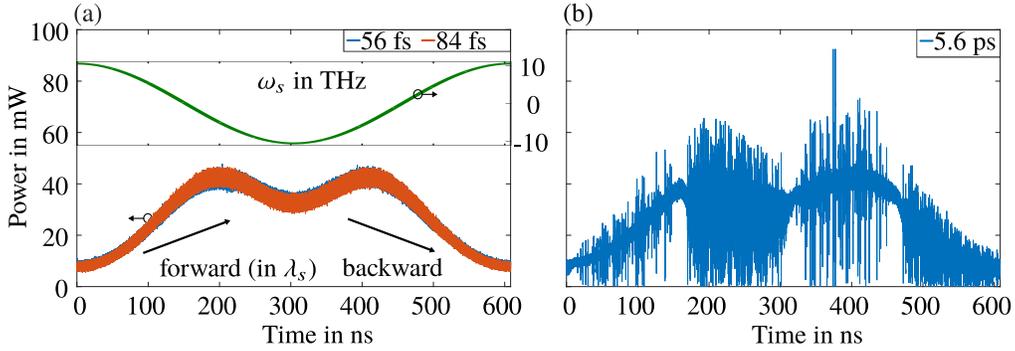}
\caption{(a) Simulated intensity trace of a forward and backward sweep in the sweet-spot operation mode with no hole formation, although dispersion is present in the fiber cavity, which is measured as the maximum GD difference due to $\tau_\mathrm{g}\left(\omega \right)$. Here $\lambda_s$ is the relative center frequency of the swept FP filter  $\omega_s$ in respect to wavelength. Note that $\omega_s = 0$ refers to the center frequency $\omega_c$ of the sweep. (b) Hole formation beyond a threshold of a finite amount of residual dispersion.}
\label{fig_threshold}
\end{figure}
\begin{figure}[t]
\centering
\includegraphics[width=\textwidth]{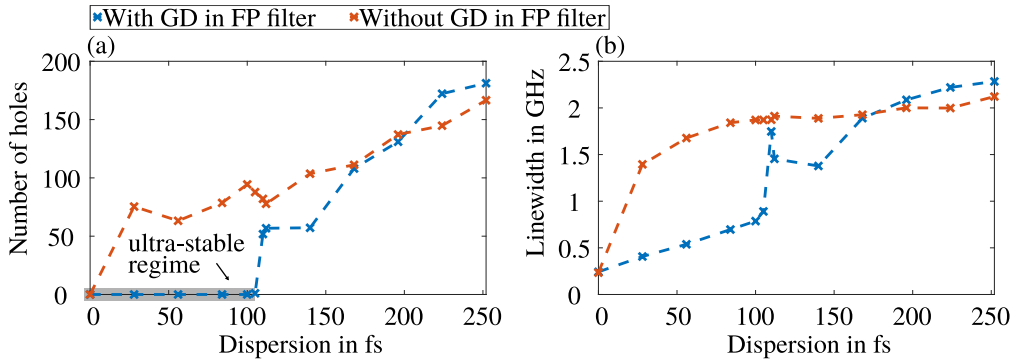}
\caption{(a) Number of holes in the intensity trace and (b) the averaged instantaneous linewidth within a roundtrip dependent on the residual dispersion in the fiber delay cavity, compared for the case with and without a GD in the FP bandpass filter.}
\label{fig_D_Comparison}
\end{figure}
\begin{figure}[t]
\centering
\includegraphics[width=\textwidth]{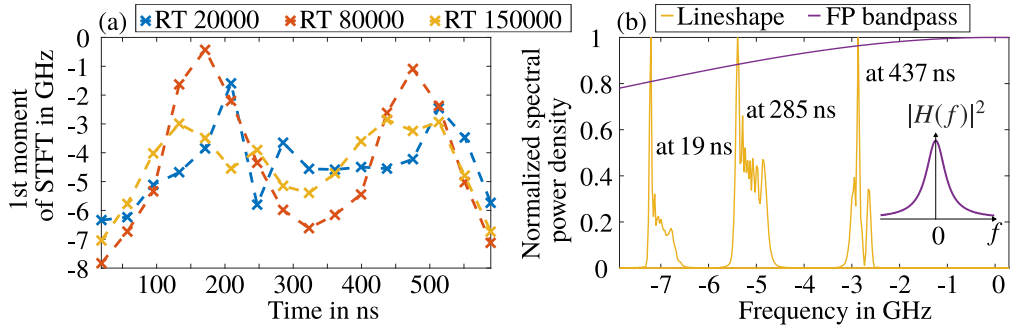}
\caption{(a) Evolution of the IF within a single roundtrip for different roundtrip numbers in the sweet-spot regime with a residual dispersion of \SI{56}{\femto\second}. (b) Instantaneous lineshapes at particular positions in the sweep of roundtrip (RT) \SI{150000}{}.}
\label{fig_OneRT_lineshape}
\end{figure}

In Fig.~\ref{fig_threshold} the intensity traces for simulations with several amounts of residual dispersion are presented for parameter values given in Table~\ref{tab_SimParam}. The parameters are selected to be close to the experimental setup described in Section \ref{sec_ExpSetup}, but with a sinusoidal sweep function and neglected switch-on and -off processes in the gain medium resulting from the buffering technique~\cite{huber2006buffered,kolb2016megahertz}. A typical value for the average power at the laser output was measured to be  $\approx \SI{30}{\milli\watt}$ and corresponds well with the value obtained in our simulations from Fig.~\ref{fig_threshold} (a), i.e. \SI{29}{\milli\watt} for a residual dispersion of \SI{56}{} or \SI{84}{\femto\second} respectively. Further information about the implementation is given in Appendix A. 
\begin{table}[t]
\caption{Simulation Parameters} 
\centering 
\begin{tabular}{|c | c | c|} 
\hline
\bf Parameter & \bf Symbol & \bf Value\\ [0.5ex] 
\hline\hline 
Fiber length & $L_\mathrm{f}$ & \SI{443.4}{\meter}  \\  \hline
Fiber group refractive index & $n_\mathrm{g}$ & \SI{1.645}{}  \\ \hline
Second-order dispersion parameter & $\beta_2 $ & \SI{9.42e-27}{\square\second\per\meter} \\\hline
Third-order dispersion parameter & $\beta_3 $ & \SI{7.62e-41}{\second\cubed\per\meter} \\ \hline
Fourth-order dispersion parameter & $\beta_4 $ & \SI{1.70e-55}{\second\tothe{4}\per\meter} \\ \hline
Fiber nonlinearity parameter &$\gamma$& \SI{2.67e-3}{\per\watt\per\meter}\\ 
\hline
Power loss in fiber spool & $\kappa_\mathrm{f}$ &\SI{0.23}{} \\ 
\hline
Reflectivity of cFBG &$R$&\SI{0.35}{}
\\ \hline
Dispersion coefficients of cFBG &$\hat{\beta}_2, \hat{\beta}_3, \hat{\beta}_4 $& variable 
\\ \hline
Center sweep frequency &  $\omega_c $& $2\pi \cdot $\SI{232.04}{\tera\hertz} (\SI{1292}{\nano\meter}) \\ \hline
Sweep bandwidth & $D_\omega $ &$2\pi \cdot$\SI{21.06}{\tera\hertz}  (\SI{117}{\nano\meter}) \\  \hline
Sweep frequency &  $\omega_0 $&  $2\pi \cdot$\SI{411}{\kilo\hertz} \\ \hline
Linewidth enhancement factor&$\alpha$&\SI{1.55}{}~\cite{cassioli2000time}\\ \hline
Carrier lifetime&$\tau_c$&\SI{70}{\pico\second}~\cite{cassioli2000time}\\\hline
Simulation bandwidth&$\Delta_\mathrm{sim}$&\SI{3.45}{\tera\hertz}\\\hline
Effective gain&$G_\mathrm{eff}$&\SI{2.06e16}{ \per\second}\\\hline
ASE noise power in simulation bandwidth&$P_\mathrm{n}$&\SI{9.05}{\milli\watt}\\\hline
FP filter bandwidth&$\Delta_\omega$& $2\pi\cdot$\SI{29.65}{\giga\hertz}  (\SI{0.165}{\nano\meter})\\ \hline
FP filter transmission&$T_\mathrm{max}$ &\SI{0.33}{}\\ \hline
Total cavity losses&$L$&\SI{15.76}{\decibel}
\\ 
\hline 
\end{tabular}
\label{tab_SimParam}
\end{table}
The sweet-spot operation mode with a residual fiber dispersion is confirmed in Fig.~\ref{fig_threshold} (a) with an ASE noise-limited intensity trace, whereas in Fig.~\ref{fig_threshold} (b) the intensity trace is distorted by high-frequency noise. This noise is dominated by consecutive holes and occurs after a threshold of residual dispersion in the fiber system is exceeded. The shape of the holes is described in detail in Section \ref{sec_SimExp}. The maximum group delay between the fastest and slowest wavelength caused by the fiber in each roundtrip is given below by the difference between the maximum and minimum value of the fiber cavity group delay $\tau_\mathrm{g}\left(\omega \right) = L_\mathrm{f} \left(\beta_2\omega + \beta_3\omega^2/2 + \beta_4\omega^3/6 \right)/c_\mathrm{scale} $ and used to quantify the residual dispersion in the fiber cavity.
\\ 
The threshold after which holes are formed in the intensity trace occurs for the system under investigation at around \SI{100}{\femto\second} of maximum group delay difference in the fiber within a \SI{117}{\nano\meter} bandwidth.\\ 
This phenomenon is demonstrated in Fig.~\ref{fig_D_Comparison} (a) which relates the number of holes in the intensity trace to the residual fiber dispersion, and shows the effect of the GD in the FP bandpass filter. Clearly, the presence of the GD in the FP bandpass filter enables hole-free operation up to a fiber dispersion of around \SI{100}{\femto\second}. The GD in the FP bandpass filter was removed by ignoring the phase shift in the frequency domain, i.e. using $|H(\omega)|$ in Eq.~(\ref{eqn_FP_update}). In addition, the average linewidth within a roundtrip in Fig.~\ref{fig_D_Comparison} (b) is reduced by more than a factor of two when the GD in the FP bandpass filter is present. 
For large residual dispersion values and a high-frequency distorted intensity trace, the linewidth as well as the number of holes of the system with GD is larger than for the system without GD, indicating a negative effect on the stability due to the GD introduced by the FP bandpass filter in this regime.
\\
The results of Fig.~\ref{fig_D_Comparison} prove that a self-stabilizing effect regulates the phase accumulation of the optical field due to the fiber dispersion in the cavity, unless a threshold is exceeded. According to our simulation, a passive mechanism, which identified here as being the frequency-dependent group delay in the FP bandpass filter, plays a major role in this process and is an indispensable requirement for hole-free operation in the sweet-spot regime. A more detailed discussion can be found in Section \ref{sec_DiscSelf}. In addition, this self-compensation makes the sweet-spot regime possible in the first place, since perfect synchronization over all wavelengths cannot be achieved in experiment.  
\\
The evolution of the IF and the instantaneous lineshape in the sweet-spot operation mode within a single roundtrip is demonstrated in Fig.~\ref{fig_OneRT_lineshape} at the output of the SOA. The position of the IF inside the static bandpass filter window is mainly determined by the dispersion perturbation and is a measure for the delay with regard to the center of the FP bandpass filter. In Fig.~\ref{fig_OneRT_lineshape} (a) the IF is extracted for different roundtrip numbers, and selected lineshapes are presented in Fig.~\ref{fig_OneRT_lineshape} (b).
The lineshapes correspond to the minimum and maximum peak values in Fig.~\ref{fig_OneRT_lineshape} (a) and to the middle of the sweep, i.e. the center window of the STFT, of roundtrip \SI{150000}{}.  
In the case presented in Fig.~\ref{fig_OneRT_lineshape}, a maximum group delay difference of \SI{56}{\femto\second} is present in the fiber delay cavity as in the case of Fig.~\ref{fig_threshold} (a).
\begin{figure}[t]
\centering
\includegraphics[width=\textwidth]{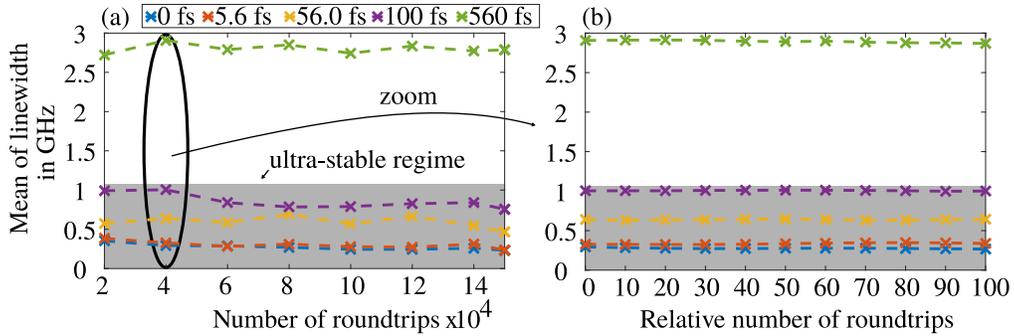}
\caption{(a) Long-term evolution of the averaged linewidth within one roundtrip for different amounts of residual dispersion. (b) Short-term evolution of the linewidth showing stationary operation, starting at roundtrip \SI{}{40000} sampled at every \SI{}{10}\textsuperscript{th} roundtrip.}
\label{fig_MeanLinewidth_Comparison}
\end{figure}
The variation of the IF is caused by the fiber dispersion which imposes a frequency chirp according to Eq.~(\ref{eqn_fiber}) on the optical field in each roundtrip. As a comparison, the IF evolution in the case of zero-dispersion in the fiber cavity exhibits a nearly straight line (see also Section~\ref{sec_DiscSelf}) instead of the oscillatory behavior in Fig.~\ref{fig_OneRT_lineshape} (a) resulting from the interplay between the FP sweep filter, the SOA gain medium and the fiber dispersion. The drift of the IF over several \SI{1000}{} roundtrips from Fig.~\ref{fig_OneRT_lineshape} (a) is stable within a tolerance of a few \SI{}{\giga\hertz}, but never converges to a constant value. When looking at the long-term evolution of the averaged instantaneous linewidth over a single roundtrip in Fig.~\ref{fig_MeanLinewidth_Comparison} (a) a similar drift can be observed over many \SI{10000}{} roundtrips, yet the short-term evolution over \SI{100}{} roundtrips in Fig.~\ref{fig_MeanLinewidth_Comparison} (b) shows a steady-state behavior. The same can be observed for the IF. The long-term variations can be addressed to the nonlinear nature of the FDML laser, and we have verified in our simulations that the stochastic influence of the ASE noise is amplifying, but does not cause the long-term drift. This long-term drift is supported by experimental data acquired in FDML-based OCT systems where coherent averaging can only be applied up to roughly \SI{100} roundtrips~\cite{pfeiffer2016flexible}, and has also been observed in previous long-term simulations \cite{jirauschek2017efficient}.  
Apparently, a stationary field does not exist, but the variations of the IF stay within a few \SI{}{\giga\hertz}. 
\subsection{Detuning tolerance}
When considering a dispersion-less cavity, i.e. $\beta_{2,3,4},\hat{\beta}_{2,3,4} = 0$, and a non-perfectly synchronized sweep filter, our simulations indicate that a detuning of up to \SI{\pm 1.0}{\milli\hertz} - equivalent to a change in the roundtrip time of \SI{\mp 5.9}{\femto\second} from the ideal resonance frequency of \SI{411}{\kilo\hertz} - does not cause hole formation in the intensity trace. The compensation mechanism remains the same as discussed in the previous section. This result is consistent with measured data where the laser was detuned from the experimentally obtained sweet-spot operation mode. Here the experimental setup was again buffered~\cite{huber2006buffered,kolb2016megahertz} and a single optical sweep was contained in \SI{300}{\nano\second}. Asymmetries between the forward and backward sweep have been observed, but are small when enforcing hole-free operation in a dispersion-free setup. The presented value here for the detuning from the ideal sweep frequency is a limit for the backward and forward sweep. 
\\In the swept-filter reference frame, a delayed optical field corresponds to a phase modulation according to
\begin{equation} \label{eqn_Delay}
u_\mathrm{out}\left(\tau\right) = u_\mathrm{in}\left(\tau\right)\exp\left[\mathrm{i} \tau_\mathrm{d}\omega_s\left(\tau\right)\right],
\end{equation} 
where $\tau_\mathrm{d}$ is the additional delay to the dispersionless roundtrip time and higher-order terms in the temporal derivatives have been neglected, as discussed in Appendix C. Thus, a delay accumulates to the phase of the optical field linearly in $\omega_s(\tau)$ while the dispersion perturbation accumulates quadratically or cubically in $\omega_s(\tau)$ as can be seen in Eq.~(\ref{eqn_Delay}) and Eq.~(\ref{eqn_fiber}).
     
\subsection{Discussion of the self-stabilization effect} \label{sec_DiscSelf}
\begin{figure}[t]
\centering
\includegraphics[width=\textwidth]{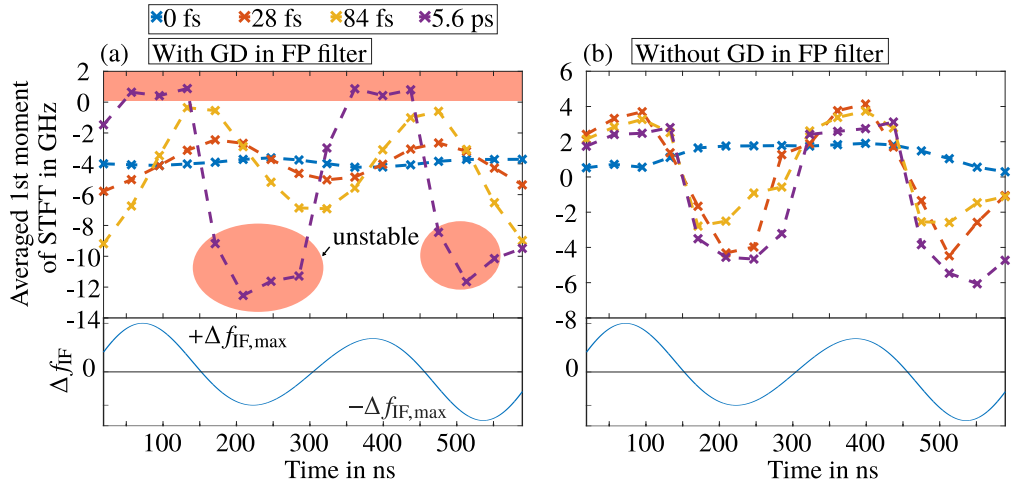}
\caption{(a) IF evolution for different amounts of residual dispersion at the output of the SOA. (b) The same as in (a) but with no GD in FP bandpass filter.}
\label{fig_IF_D_Comparison}
\end{figure}

\begin{figure}[t]
\centering
\includegraphics[width=\textwidth]{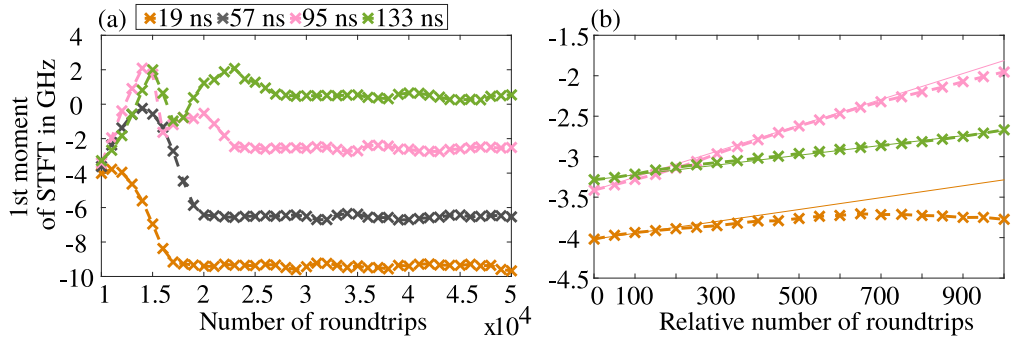}
\caption{(a) Long-term drift of the IF at specific points in time or equivalently of single wavelengths over \SI{40000}{} roundtrips when switching on a fiber dispersion of \SI{100}{\femto\second } after \SI{10000}{} roundtrips until the quasi-steady-state is reached. (b) Zoom into the first 1000 roundtrips compared to the dispersion accumulation in the fiber
(solid line).} 
\label{fig_DSwitchON_IF_RT_evolution}
\end{figure}

The fiber dispersion and the detuning from the ideal filter frequency introduce timing delays in the optical field relative to the moving transmission window of the swept bandpass filter, i.e. cause a frequency-dependent roundtrip time $RTT(\omega)$.\\
One quantity to measure the introduced time delays is the IF of the optical field, which describes the offset from the center of the static bandpass filter in the swept-filter reference frame as mentioned above. The shift in the IF caused by the fiber dispersion or by a detuning can be obtained by differentiating the phase of the complex exponential in Eq.~(\ref{eqn_fiber}) or Eq.~(\ref{eqn_Delay}) with respect to time. For example, in the case of the fiber dispersion, this yields 
\begin{equation}  \label{eqn_fiber_IFshift}
\Delta f_{\mathrm{IF}} = -1/(2\pi) L_\mathrm{f}\omega_s(\tau) \frac{\partial \omega_s(\tau)}{\partial \tau }\left[\beta_2 + \frac{1}{2}\beta_3\omega_s(\tau) + \frac{1}{6}\beta_4 \omega_s^2(\tau) \right]  
\end{equation} 
per pass through the fiber spool.
\\
An analytic treatment of the self-stabilization mechanism is tedious due to the nonlinear dependencies such as the frequency-dependent GD of the FP filter with respect to frequency, and the complex interplay of the processes in the SOA gain medium, the filter and the fiber over a long time scale. Thus, we focus on an intuitive explanation in the following.  
\\
The frequency-dependent GD of the bandpass filter $GD(\omega)$ is able to reduce the roundtrip time for a specific IF when moved away from the center of the filter, where the GD is maximum for a Lorentzian filter shape. The shape of $GD(\omega)$ of the static bandpass filter in the frequency domain is given by 
\begin{equation}
 GD(\omega) = \frac{2}{\Delta_\omega}\frac{1}{1 + \left(2\omega / \Delta_\omega \right)^2}.
\end{equation}
At $\omega = 0$, in the center of the bandpass filter, the GD is \SI{10.7}{\pico\second} and always less when $|\omega| > 0$, e.g. reduced by \SI{730}{\femto\second} when shifted by \SI{-4}{\giga\hertz}. In this way, a delayed IF with respect to the center of the filter can be accelerated over consecutive roundtrips to a position in the filter window where no holes form within the threshold limit, whereas an accelerated IF would be further accelerated, leading to an unstable drift. Related further discussion can be found in~\cite{pfeiffer2018ultra}.
The selection of the stable filter position depends on the amount of residual dispersion in the fiber as discussed above and is demonstrated in Fig.~\ref{fig_IF_D_Comparison} (a). Here, it is evident that the IF chooses a different position in the filter window dependent on the fiber dispersion, in order to operate in the sweet-spot regime. In comparison, when the bandpass filter has no GD in Fig.~\ref{fig_IF_D_Comparison} (b) the IF does not differ significantly with different amounts of dispersion in the fiber cavity, and holes are formed as demonstrated in Fig.~\ref{fig_D_Comparison} (a). One important factor is here that the modified roundtrip times are not compensated for.  
For the case of \SI{5.6}{\pico\second}, the intensity trace is distorted by high-frequency noise as presented in Fig.~\ref{fig_threshold} (b), and in Fig.~\ref{fig_IF_D_Comparison} (a), the evolution of the IF follows the fiber chirp $\Delta f_\mathrm{IF}$ which is depicted below, when taking the zero-dispersion line as a reference. The same holds for the case without GD in the FP filter. Again this shows that the accumulated chirp per roundtrip due to the fiber dispersion cannot be compensated in these cases. Yet, single narrow sweet-spots exist when crossing the zero-dispersion line, see Fig.~\ref{fig_threshold} (b).
\\
Another important feature of the self-stabilization effect is the temporal extension over several thousands of roundtrips until the quasi-steady-state regime of Fig.~\ref{fig_IF_D_Comparison} (a) is reached. This can be demonstrated by starting with zero-dispersion in the fiber cavity and abruptly switching on the fiber dispersion after \SI{10000}{} roundtrips. In Fig.~\ref{fig_DSwitchON_IF_RT_evolution} (a) this scenario is shown for a residual dispersion of \SI{100}{\femto\second} where the maximum shift in the IF is $\Delta f_\mathrm{IF,max} = \pm \SI{3.6}{\mega\hertz}$ per roundtrip. Note that in the case of \SI{100}{\femto\second} the long-term drift of the IF as discussed in Fig.~\ref{fig_OneRT_lineshape} (a) for \SI{56}{\femto\second} is far less pronounced. In the first few hundred roundrips the IF follows the dispersion accumulation in the fiber as visualized in Fig.~\ref{fig_DSwitchON_IF_RT_evolution} (b) where the solid lines are obtained from $n\Delta f_\mathrm{IF}(\tau_w) $ with the roundtrip number $n$ and the temporal center of the STFT window $\tau_w$. In a single roundtrip, this shift in the IF is equivalent to a change in the roundtrip time. The difference in $GD(\omega)$ in the FP filter between the zero-dispersion position of approximately \SI{-4}{\giga\hertz} shifted by \SI{-3.6}{\mega\hertz} is $\approx\SI{1}{\femto\second}$ and thus negligible compared to the delays introduced by the dispersion. Yet, during the long-term accumulation, this difference increases to $\approx \SI{800}{\femto\second}$ when shifted by \SI{-2}{\giga\hertz}, which exceeds the maximum dispersion delay by a factor of \SI{8}{}. Hence, the frequency offset required for the compensation due to the group delay of the FP filter builds up over many roundtrips.\\
Based on our findings, it can be speculated that by engineering the GD in the FP filter by e.g. using
chirped dielectric coatings on the FP filter mirrors or modifying the filter bandwidth, the system can be made more robust against residual fiber dispersion or environmental changes. An increase in the stability threshold found in Section~\ref{sec_DispComp} could enable a widespread use of the ultra-stable operating regime, which requires up to now a high level of synchronization. The model presented here serves as an ideal starting point for such optimization problems.
\subsection{Influence of SOA parameters on the sweet-spot operation mode }
The carrier lifetime $\tau_c$, the linewidth enhancement factor $\alpha$  and the saturation power $P_\mathrm{sat}$ of the SOA gain medium play an important role in obtaining the sweet-spot operation mode. Especially the phase accumulation due to the linewidth enhancement factor in combination with the gain recovery dynamics and the interplay with the swept FP filter can create chaotic intensity traces. With typical values of $\tau_c = \SI{380}{\pico\second}$ and $\alpha = 5$ as previously used in~\cite{jirauschek2009theoretical,jirauschek2017efficient} a sweet-spot operation mode could not be achieved. An InP/InGaAsP multiple-quantum-well (MQW) booster SOA was used in~\cite{pfeiffer2018ultra} (Thorlabs BOA1132S) whose physical properties at the corresponding operating condition need to be known. Yet, to the best of our knowledge, no rigorious characterization of $\alpha$ exists in the specific wavelength range of interest for this device. It is commonly known that strained MQW structures exhibit a reduced linewidth enhancement factor due to the high differential gain compared to bulk materials~\cite{kano1993reduction,zah1994high,tiemeijer1991dependence,thiis1994progress}, and values of $\alpha \le 2$ were reported for the same wavelength and material system for lasers. Similarly, the SOA here serves as the laser gain medium, and essentially compensates for the roundtrip losses, thus operating in the saturated regime. Dynamical aspects, similar to SOAs for signal amplification, come into play for the hole formation in FDML lasers. In this case, the literature values found for $\alpha$ are far more diverse, with some references stating similar values as above \cite{dorren2004all,huang2011optimized,grigoryan2006soa} and others reporting values of \SI{3} or higher for various devices and operating conditions~\cite{zilkie2008time,mecozzi1997saturationFWM,
mecozzi1997saturation,storkfelt1991measurement,clavero2005all,tiemeijer1996self,bilenca2006numerical}. When using the parameters $\tau_c = \SI{440}{\pico\second}$, $\alpha = 3$ from a study of SOA fiber ring lasers with a static bandpass filter~\cite{girard2011soa}, a sweet-spot over the whole sweep bandwidth is found in our simulation. However, at moderate dispersion levels, fringes occur in the hole minima which cannot be observed in experiment, as will be discussed in the section below. Therefore, we conclude that the above lower $\alpha$ value, as reported for strained MQW laser operation, yields the best agreement with the shape and time scale of the holes in experiment, especially at moderate dispersion levels. Further simulations reveal that a shorter carrier lifetime combined with a higher $\alpha$ factor increases the occurrence of fringes in the holes. In~\cite{wang2007temporal}, it was discussed that a time-dependent $\alpha$-factor which also depends on the carrier density and wavelength might be more adequate for accurately modeling highly dynamic processes in the order of \SI{}{\pico\second}. However, due to the numerical load of our simulations and the intention to derive a compact model suitable for optimization purposes, this approach will be left to future work. On the other hand, carrier heating and spectral hole burning have been found to be negligible at the time scale of the holes~\cite{wang2007temporal,mecozzi1997saturation,mork1997theory}.

\section{Experimental verification} \label{sec_SimExp}
\begin{figure}[t]
\centering
\includegraphics[width=\textwidth]{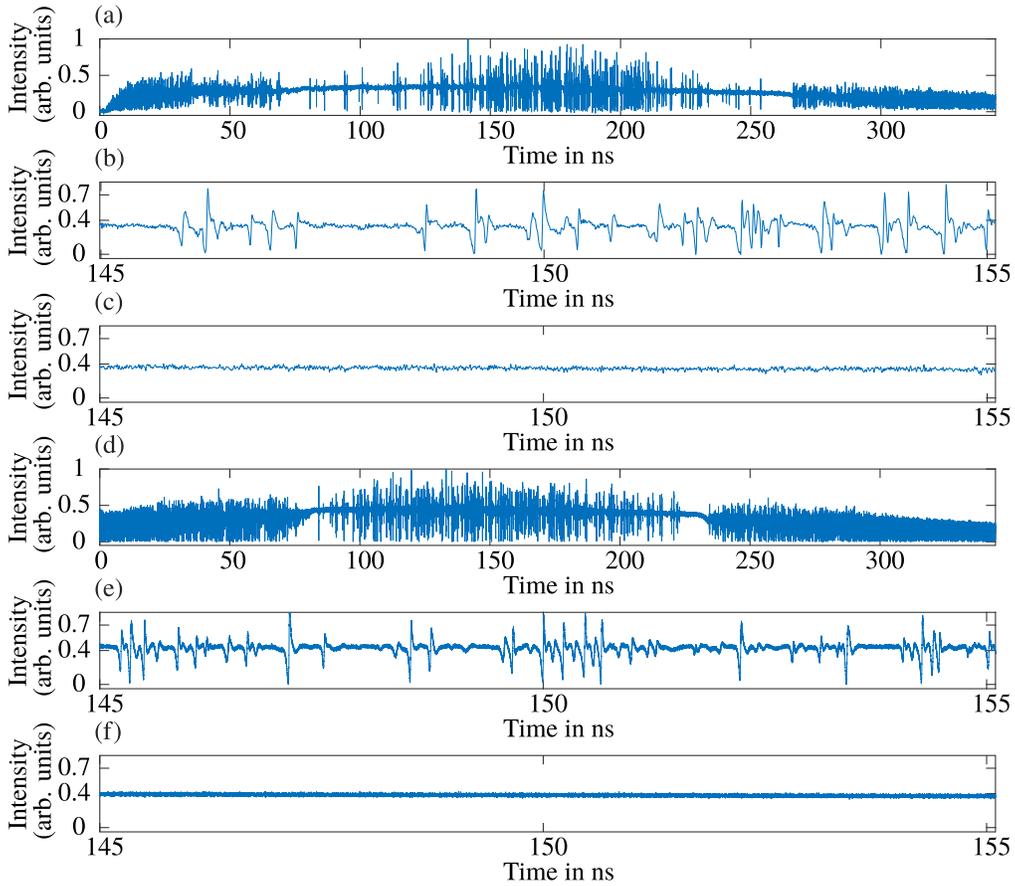}
\caption{(a) Measured intensity trace of a laser with a non-dispersion-compensated SMF-28 fiber cavity. The backward sweep of \SI{104}{\nano\meter} is centered around \SI{1307}{\nano\meter} at a filter sweep frequency of \SI{363}{\kilo\hertz}. (b) Zoom into a noisy region in the intensity trace, showing the hole dynamics. (c) Identical zoom into the experimentally obtained sweet-spot. (d) A simulated intensity trace of the case in (a) and (e) zoom into noisy region for comparison. (f) Zoom into the sweet-spot with a residual dispersion of \SI{84}{\femto\second}. }
\label{fig_SimExp_comparison}
\end{figure}

\begin{figure}[t]
\centering
\includegraphics[width = \textwidth]{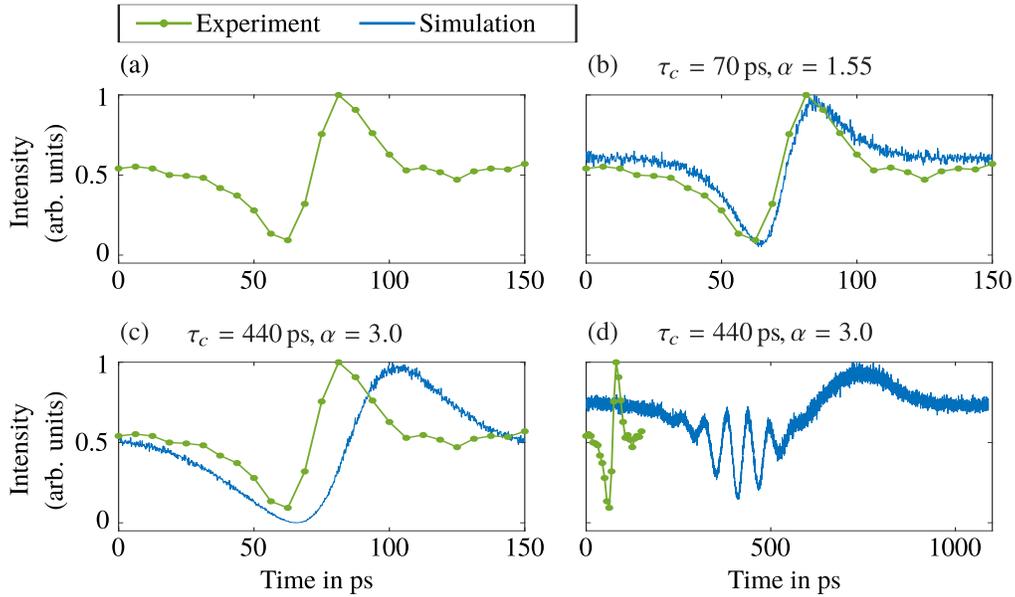}
\caption{(a) Measured hole in a laser setup which was detuned by \SI{-100}{\milli\hertz} from the ultra-stable regime. (b) A hole extracted from simulation with the SOA parameters used in this work, see Table~\ref{tab_SimParam}. (c) Hole for the SOA parameters from~\cite{girard2011soa}, yet with a longer duration than in the experiment. (d) Fringe-distorted hole for the SOA parameters in (c) which could not be observed in experimentally measured intensity traces.}
\label{fig_Hole_Comparison}
\end{figure}

In order to verify our physical model, we reproduced a detuned intensity trace of an FDML laser with a non-dispersion-compensated SMF-28 fiber featuring the full noise dynamics due to the dispersion perturbation. The fiber parameters and the dispersion coefficients were extracted from the manufacturer specification~\cite{SMF28}. The measured intensity trace of a backward sweep of \SI{104}{\nano\meter} centered around \SI{1307}{\nano\meter} is shown in Fig.~\ref{fig_SimExp_comparison} (a). 
The bandpass filter was driven with a frequency of \SI{363}{\kilo\hertz} which was slightly detuned from the sweep frequency where the center wavelength matches the roundtrip time in the laser cavity, resulting in a noisy intensity trace with a single narrow sweet-spot around the middle of the sweep. In the detuned case, two narrow sweet-spots can be found in the intensity trace, in Fig~\ref{fig_SimExp_comparison} (a) at around \SI{75}{\nano\second} and \SI{260}{\nano\second}, because of the parabolically shaped dispersion relation which causes two wavelengths to be synchronized to the filter sweep
rate. The measurement was obtained with a real-time oscilloscope and a high-speed photodiode, with a total analog bandwidth of \SI{50}{\giga\hertz}.\\  
The hole dynamics can be observed in Fig.~\ref{fig_SimExp_comparison} (b) where a \SI{10}{\nano\second} section around \SI{150}{\nano\second} is cut out. The characteristic shape of the holes can be identified as a power drop of up to \SI{100}{\percent} followed by an overshoot caused by the gain-recovery dynamics in the SOA gain medium. 
The duration of the holes is, among other things, influenced by the carrier lifetime $\tau_c$ of the SOA,  as well as the linewidth enhancement factor $\alpha$, and is in the range of \SIrange{50}{70}{\pico\second}.\\ 
The reproduced intensity trace, i.e. $|u(\tau)|^2 $, at the laser output after \SI{40000}{} roundtrips is shown in Figs.~\ref{fig_SimExp_comparison} (d) and (e) respectively. Because of the quasi-stationary nature of the FDML laser, the exact position and shape of the holes can barely be captured in the simulation. Thus, the focus lies here in a qualitative agreement, where we pay special attention to the characteristics of the individual holes. The laser was detuned by \SI{-0.8}{\hertz} to obtain the closest match. All in all, good agreement between the simulated and experimental traces can be observed with regard to the hole densities and shapes. Since the switch-on and -off process in the SOA gain medium in the buffering mode is not considered in the simulation, the left and right edges up to \SI{20}{\nano\second} cannot be compared. This effect has no influence on the physical mechanism behind the self-stabilization, and is therefore neglected.\\ 
Figures~\ref{fig_Hole_Comparison} (a) and (b) demonstrate that single holes can be precisely reproduced with the chosen SOA parameters from Table~\ref{tab_SimParam}. 
For $\tau_c = \SI{440}{\pico\second}$ and $\alpha = 3$~\cite{girard2011soa} the hole duration increases by a factor of \SI{2}{}, see Fig.~\ref{fig_Hole_Comparison} (c). Furthermore, in intensity traces with only few holes, special patterns emerge with up to a factor of \SI{10}{} increased hole durations and fringe formation in the minima, see Fig.~\ref{fig_Hole_Comparison} (d). Such a feature has not been observed in any measured intensity trace. This indicates the sensitivity of the hole dynamics on the SOA parameters and confirms that a carrier lifetime of $\approx \SI{100}{\pico\second}$ along with a linewidth enhancement factor $\alpha \approx 1.5$ are a reasonable choice in our case. This is of particular importance, since due to the strong phase-amplitude coupling caused by the narrow sweep filter, a hole contains information about the interplay of all cavity elements as well as the gain recovery dynamics which is visible in the overshoot. The number of holes in the intensity trace was compared in a region between the two narrow sweet-spots from \SI{80}{\nano\second} to \SI{230}{\nano\second} where the hole density is moderate. \SI{356}{} and \SI{315}{} holes were counted in experiment and in the simulation respectively, which corresponds to a deviation of \SI{12}{\percent} and is plausible when considering the above-mentioned uncertainties.
\\
The good agreement between simulation and experiment shows that all relevant physical mechanisms are contained in our model. Neglecting polarization effects, such as bending-induced birefringence or polarization-mode dispersion in the optical fiber, appears to be a legitimate assumption for a single-polarization SOA, where only a single-polarization component experiences considerable gain. In experiment, a polarization controller before the SOA is used in order to adjust the polarization state of the light field to the axis where the polarization-dependent SOA has maximum gain, and to minimize timing delays introduced by the polarization dynamics. Previous investigations into the polarization state of FDML lasers with a single-polarization SOA operating far beyond the sweet-spot regime showed a well-defined polarization state over the whole sweep bandwidth~\cite{wieser2012chromatic}. This also agrees with the observation in experiment in the sweet-spot regime that the polarization state of the light field reproduces after each roundtrip at the entrance of the SOA, and does not significantly vary or cause instabilities. \\
In the case of the numerical simulations described in Section \ref{sec_DispComp}, the SMF-28 fiber was replaced by the \SI{1}{}:\,\SI{8}{} mix of a SMF-28 and a HI-1060~\cite{HI1060} fiber as discussed above. For both fiber cavities, Turing-like instabilities which are discussed in~\cite{slepneva2013dynamics} were observed in parts of the intensity trace in the case of detuning or residual dispersion, along with oscillatory and irregular patterns. This indicates that the hole dynamics can differ in particular cases, typically near the turning point of the sweep filter waveform, in view of the forms seen in Figs.~\ref{fig_SimExp_comparison} (b) and (e). In general, the occurrence of intensity patterns is not stationary over consecutive roundtrips.  

\section{Conclusion}
We have demonstrated by numerical simulation of an FDML laser that a residual dispersion in the fiber cavity which delays or accelerates each instantaneous frequency of the optical field relative to the FP sweep filter can be 
compensated by the frequency-dependent group delay in the swept bandpass filter, such that no holes occur in the intensity trace of the laser.\\
Systems without this self-compensation mechanism show less coherence and exhibit a hole-distorted intensity trace because the dispersion perturbation 
accumulates within each roundtrip without compensation, and drives the system out of its stability region. For the same reason a detuning from the ideal sweep filter frequency of up to \SI{\pm 1.0}{\milli\hertz} 
can be regulated for the system of investigation with no dispersion in the optical fiber. 
Our simulations display good correspondence with experimental data, and show that in a single-polarization setup, the hole dynamics is determined 
by the complex interplay of the dispersive fiber, the swept bandpass filter and the nonlinear dynamics of the SOA gain medium.\\  
Within the sweet-spot regime, the generation of FDML-based frequency combs is highly attractive, and the temporal compression
limit of the wavelength-swept laser output for short pulse generation may be reached, which
is in the sub-\SI{}{\pico\second} range~\cite{eigenwillig2013picosecond}. This ultra-stable operating regime with its enhanced coherence properties
offers new, unexplored possibilities in sensing and imaging applications.

\section*{Appendix A: Additional implementation information}
\subsection*{Fiber}
The fiber delay line consists of a \SI{1}{}:\,\SI{8}{} mix of a SMF-28~\cite{SMF28} and a HI-1060~\cite{HI1060} fiber to shift the zero-dispersion wavelength out of the sweep window such that the cFBG can be used for dispersion compensation. Here, we used a model for the dispersion coefficient $\beta_2$ of the SMF-28 fiber provided by the manufacturer~\cite{SMF28} which was fitted to the experimentally obtained dispersion-relation of the fiber mix. Thus, adapted values were used for the fiber length (\SI{443.4}{\meter}), the zero-dispersion wavelength (\SI{1418.9}{\nano\meter}) and the zero-dispersion slope (\SI{0.07}{\pico\second\per\nano\meter\squared\per\kilo\meter}) which are required for the manufacturer's model. In addition, this model was verified experimentally.

\subsection*{Buffering}
The experimental laser curve was recorded in the buffered mode~\cite{huber2006buffered,kolb2016megahertz} when the laser is run in a \SI{1/8}{} duty cycle in order to increase the sweep frequency by copying single sweeps of the optical field, after out-coupling the light from the cavity. In this way, the sweep rate can be increased by increasing the amplitude of the FP sweep filter drive function without increasing its mechanical resonance frequency, which constitutes a bottleneck for the maximum FDML sweep rate~\cite{wieser20105MHzOCT}. In the simulation, the off-time of 7/8 of the sweep was not considered. Instead, we used the physical fiber length of \SI{443.4}{\meter} (\SI{562.6}{\meter} in Section~\ref{sec_SimExp}) in Eq.~(\ref{eqn_fiber}) and replaced the corresponding sweep rate of \SI{411}{\kilo\hertz} (\SI{363}{\kilo\hertz}) by $8\cdot\SI{411}{\kilo\hertz}$ ($8\cdot\SI{363}{\kilo\hertz}$) for a single sweep. In doing so, we obtained a continuous sweep with the correct sweep timings and the actual physical propagation length in the cavity. However, we neglected the switch-on and -off process in the SOA due to the continuous sweep, as well as the off-time, which carries no relevant physical information. The good agreement in Section~\ref{sec_SimExp} validates this approach.\\
Because of the buffering technique, the sweep of the bandpass filter is nearly linear since only at \SI{1/8}{} of the sweep time around the zero crossing point of the sinusoidal sweep function is the gain medium switched on. A linear sweep function was therefore used in Section~\ref{sec_SimExp}.   

\subsection*{SOA}
Equation~(\ref{eqn_SOA_ODE}) was numerically integrated with a fourth-order Runge-Kutta method. The unsaturated gain was interpolated with a Gaussian and the saturation power with a fourth-order polynomial. Note that the measured input saturation power $P_\mathrm{sat,m,in}$ is related with $P_\mathrm{sat}$ in Eq.~(\ref{eqn_SOA_ODE}) by $P_\mathrm{sat} = \left[\exp\left(h_0\right) -  2\right]/\left[ 2\log(2) \right] P_\mathrm{sat,m,in}$. 

\section*{Appendix B: Characterization of the ultra-stable operating regime} 
In order to quantify the coherence properties of the ultra-stable operating regime, we determine the width of the instantaneous lineshape and the number of local minima of the high-frequency fluctuations in the intensity trace, i.e the number of holes. The instantaneous lineshape is a local measure of the phase evolution and of fluctuations in the optical field, whereas the high-frequency noise in the intensity trace is a result of the complex laser dynamics evolving over many roundtrips. In addition, we describe the optical field by its IF, which measures the offset from the center frequency of the bandpass filter in the swept-filter reference frame. The three above-mentioned quantities are especially important for the imaging properties of FDML lasers~\cite{pfeiffer2018ultra}.\\
We extract the instantaneous linewidth and frequency by performing a short-time Fourier transformation (STFT) of the complex field envelope, where we use a rectangular window function and divide a single sweep into 16 pieces. For the parameter values chosen here, this corresponds to a minimum theoretical time-frequency uncertainty of \SI{2.09}{\mega\hertz} or \SI{38.02}{\nano\second} respectively for $T =\SI{608.27}{\nano\second} $ when neglecting the broadening due to the rectangular window. A similar approach was followed in \cite{todor2011instantaneous,todor2012balance} where it proved possible to reproduce experimental measurements of the instantaneous laser lineshape. The IF is estimated by the first moment within the STFT window, and the linewidth is estimated by the root mean square bandwidth (RMS) around the mean frequency. The RMS is used here since the lineshape can have different appearances with e.g. a noisy multi-peak structure where a full width at half maximum (FWHM) approach would be inappropriate, and was also used in~\cite{girard2011soa} to characterize SOA fiber ring lasers with a static bandpass filter which are similar to FDML lasers. The blurring effect of noise on the RMS bandwidth was reduced here by using a rectangular window of \SI{30}{\giga\hertz} centered around the peak frequency of the lineshape which is sharply concentrated around the IF, as can be seen in Fig.~\ref{fig_OneRT_lineshape} (b). \\
The number of holes is determined by counting the local minima exceeding a threshold of \SI{30}{\percent} from the mean of the intensity trace, obtained by applying a Gaussian low-pass filter with a FWHM width of \SI{100}{\mega\hertz}. To be robust against noise, the minimum peak distance was chosen to be \SI{40}{\pico\second} and the minimum peak width was set to \SI{5}{\pico\second}. Furthermore, we limited the bandwidth of the simulated intensity traces with a Gaussian low-pass filter with a FWHM width of \SI{500}{\giga\hertz}, in order to reduce the influence of the ASE noise on the hole counting algorithm.\\
The IF in Fig.~\ref{fig_IF_D_Comparison} at each computed point was averaged over a period from roundtrips \SI{40000}{} to \SI{150000}{}, sampled every $\SI{20000}{}$\textsuperscript{th} roundtrip in order to smooth the drift of the IF which was discussed in Fig.~\ref{fig_OneRT_lineshape} (a). The number of holes in Fig.~\ref{fig_D_Comparison} (a) as well as the mean linewidth per roundtrip in Fig.~\ref{fig_D_Comparison} (b) were averaged in the same way, though starting at roundtrip \SI{100000}{}, since we observed that certain low amplitude patterns in the intensity trace get smoothed out over this long time scale.
\section*{Appendix C: Delay in the swept-filter reference frame}
The transformation of the optical field envelope $A\left(\tau\right)$ with the retarded time $\tau$ to the swept-filter reference frame envelope $u\left(\tau\right)$~\cite{jirauschek2009theoretical} is performed by
\begin{equation}
 u(\tau)=A(\tau) \exp\left[\mathrm{i}\varphi(\tau)\right].
\end{equation}
Here, 
\begin{equation}
\varphi(\tau) = \int \left[ \Omega(\tau^\prime) - \omega_c \right] \mathrm{d}\tau^\prime
\end{equation}
which modulates the IF of the optical field in accordance with the center frequency of the optical bandpass filter $\Omega(\tau^\prime)$, shifted by the center frequency of the sweep $\omega_c$. In the case when the optical field is delayed by a time $\tau_\mathrm{d} $ the delayed field in the swept-filter reference frame is
\begin{equation}
u_\mathrm{d}(\tau) = A(\tau - \tau_\mathrm{d})\exp\left[\mathrm{i}\varphi(\tau)\right].
\end{equation}
The delayed field envelope can be rewritten in terms of $A(\tau)$ as
\begin{equation} \label{eqn_DelayOperator}
A(\tau - \tau_\mathrm{d}) = \sum_{k = 0}^{\infty}\frac{\left(-\tau_\mathrm{d}\frac{\partial}{\partial \tau}\right)^k}{k!}A(\tau),
\end{equation}
which follows from the Fourier transform properties, and is known as the time-delay operator $\exp\left(-\tau_\mathrm{d}\frac{\partial}{\partial \tau} \right)$.
When using the property of Eq.~(\ref{eqn_DelayOperator}) and neglecting higher order time derivatives of $A\left(\tau \right)$ as in~\cite{jirauschek2009theoretical,jirauschek2017efficient} according to
\begin{equation}
 \left( \frac{\partial}{\partial \tau} \right)^j  A(\tau) \approx \left[ -\mathrm{i} \omega_s(\tau) \right]^j  u(\tau)\exp\left[-\mathrm{i}\varphi(\tau)\right]
\end{equation}
 with $\omega_s(\tau) = \Omega(\tau) - \omega_c  $, we obtain
\setlength{\arraycolsep}{0.0em}
\begin{eqnarray}
u_\mathrm{d}(\tau) \, && = \left[\sum_{k = 0}^{\infty}\frac{\left(-\tau_\mathrm{d}\frac{\partial}{\partial \tau}\right)^k}{k!}A(\tau) \right]\exp\left[\mathrm{i}\varphi(\tau)\right]
\\&&= \left[ A(\tau) + \left(-\tau_\mathrm{d}\right) \frac{\partial}{\partial \tau}  A(\tau) + \frac{\left(-\tau_\mathrm{d}\right)^2}{2!} \frac{\partial^2}{\partial \tau^2 }  A(\tau) + ...\right] \exp\left[\mathrm{i}\varphi(\tau)\right]
\\
 &&= u(\tau) + \left(-\tau_\mathrm{d}\right) \left[ -\mathrm{i}\omega_s(\tau) \right]  u(\tau) + \frac{\left(-\tau_\mathrm{d}\right)^2}{2!}\left[ -\mathrm{i}\omega_s(\tau) \right]^2  u(\tau) + ...
\\
 &&= u(\tau)\left[1   -\mathrm{i} \left(-\tau_\mathrm{d}\right)\omega_s(\tau)  +  \frac{\left[ -\mathrm{i}\left(-\tau_\mathrm{d}\right)\omega_s(\tau) \right]^2}{2!}  + ... \right]
\\
 &&= u(\tau)\exp\left[ -\mathrm{i}\left(-\tau_\mathrm{d}\right)\omega_s(\tau) \right].
\end{eqnarray}
\setlength{\arraycolsep}{5pt}

\section*{Funding}
Deutsche Forschungsgemeinschaft (JI 115/4-2, JI 115/8-1, HU 1006/6); H2020 European Research Council (CoG no. 646669); Bundesministerium f\"ur Bildung und Forschung (13GW0227B).

\section*{Disclosures}
TP: Optores GmbH (E,P,R), RH: Optores GmbH (I,P,R), Optovue Inc. (P,R), Zeiss Meditec (P,R), Abott (P,R).

\bibliography{references}

\begin{thebibliography}{10}
\newcommand{\enquote}[1]{``#1''}

\bibitem{jirauschek2015wavelength}
C.~Jirauschek and R.~Huber, \enquote{Wavelength shifting of intra-cavity
  photons: Adiabatic wavelength tuning in rapidly wavelength-swept lasers,}
  {\protect\JournalTitle{Biomedical Optics Express}} \textbf{6}, 2448--2465
  (2015).

\bibitem{SiegmanLasers}
A.~E. Siegman, \emph{Lasers} (University Science Books, 1986).

\bibitem{huang1991optical}
D.~Huang, E.~A. Swanson, C.~P. Lin, J.~S. Schuman, W.~G. Stinson, W.~Chang,
  M.~R. Hee, T.~Flotte, K.~Gregory, C.~A. Puliafito, and J.~G. Fujimoto,
  \enquote{Optical coherence tomography,} {\protect\JournalTitle{Science}}
  \textbf{254}, 1178--1181 (1991).

\bibitem{wieser20105MHzOCT}
W.~Wieser, B.~R. Biedermann, T.~Klein, C.~M. Eigenwillig, and R.~Huber,
  \enquote{Multi-megahertz {OCT}: High quality 3{D} imaging at 20 million
  {A}-scans and 4.5 {GV}oxels per second,} {\protect\JournalTitle{Optics
  Express}} \textbf{18}, 14685--14704 (2010).

\bibitem{biedermann2008real}
B.~R. Biedermann, W.~Wieser, C.~M. Eigenwillig, G.~Palte, D.~C. Adler, V.~J.
  Srinivasan, J.~G. Fujimoto, and R.~Huber, \enquote{Real time en face
  {F}ourier-domain optical coherence tomography with direct hardware frequency
  demodulation,} {\protect\JournalTitle{Optics Letters}} \textbf{33},
  2556--2558 (2008).

\bibitem{adler2011extended}
D.~C. Adler, W.~Wieser, F.~Trepanier, J.~M. Schmitt, and R.~A. Huber,
  \enquote{Extended coherence length {F}ourier domain mode locked lasers at
  1310 nm,} {\protect\JournalTitle{Optics Express}} \textbf{19}, 20930--20939
  (2011).

\bibitem{reznicek2014megahertz}
L.~Reznicek, T.~Klein, W.~Wieser, M.~Kernt, A.~Wolf, C.~Haritoglou, A.~Kampik,
  R.~Huber, and A.~S. Neubauer, \enquote{Megahertz ultra-wide-field
  swept-source retina optical coherence tomography compared to current existing
  imaging devices,} {\protect\JournalTitle{Graefe's Archive for Clinical and
  Experimental Ophthalmology}} \textbf{252}, 1009--1016 (2014).

\bibitem{klein2013joint}
T.~Klein, R.~Andr{\'e}, W.~Wieser, T.~Pfeiffer, and R.~Huber, \enquote{Joint
  aperture detection for speckle reduction and increased collection efficiency
  in ophthalmic {MH}z {OCT},} {\protect\JournalTitle{Biomedical Optics
  Express}} \textbf{4}, 619--634 (2013).

\bibitem{biedermann2009recent}
B.~R. Biedermann, W.~Wieser, C.~M. Eigenwillig, and R.~Huber, \enquote{Recent
  developments in {F}ourier {D}omain {M}ode {L}ocked lasers for optical
  coherence tomography: {I}maging at 1310 nm vs. 1550 nm wavelength,}
  {\protect\JournalTitle{Journal of Biophotonics}} \textbf{2}, 357--363 (2009).

\bibitem{jung2008characterization}
E.~J. Jung, C.-S. Kim, M.~Y. Jeong, M.~K. Kim, M.~Y. Jeon, W.~Jung, and
  Z.~Chen, \enquote{Characterization of {F}{B}{G} sensor interrogation based on
  a {F}{D}{M}{L} wavelength swept laser,} {\protect\JournalTitle{Optics
  Express}} \textbf{16}, 16552--16560 (2008).

\bibitem{lee2011remote}
B.~C. Lee and M.~Y. Jeon, \enquote{Remote fiber sensor based on cascaded
  {F}ourier domain mode-locked laser,} {\protect\JournalTitle{Optics
  Communications}} \textbf{284}, 4607--4610 (2011).

\bibitem{lee2010dynamic}
B.~C. Lee, E.-J. Jung, C.-S. Kim, and M.~Y. Jeon, \enquote{Dynamic and static
  strain fiber {B}ragg grating sensor interrogation with a 1.3 $\mu$m {F}ourier
  domain mode-locked wavelength-swept laser,}
  {\protect\JournalTitle{Measurement Science and Technology}} \textbf{21},
  094008 (2010).

\bibitem{kranendonk2007high}
L.~A. Kranendonk, X.~An, A.~W. Caswell, R.~E. Herold, S.~T. Sanders, R.~Huber,
  J.~G. Fujimoto, Y.~Okura, and Y.~Urata, \enquote{High speed engine gas
  thermometry by {F}ourier-domain mode-locked laser absorption spectroscopy,}
  {\protect\JournalTitle{Optics Express}} \textbf{15}, 15115--15128 (2007).

\bibitem{kranendonk2005modeless}
L.~A. Kranendonk, R.~J. Bartula, and S.~T. Sanders, \enquote{Modeless operation
  of a wavelength-agile laser by high-speed cavity length changes,}
  {\protect\JournalTitle{Optics Express}} \textbf{13}, 1498--1507 (2005).

\bibitem{kranendonk2007wavelength}
L.~A. Kranendonk, R.~Huber, J.~G. Fujimoto, and S.~T. Sanders,
  \enquote{Wavelength-agile {H}$_2${O} absorption spectrometer for thermometry
  of general combustion gases,} {\protect\JournalTitle{Proceedings of the
  Combustion Institute}} \textbf{31}, 783--790 (2007).

\bibitem{karpf2015time}
S.~Karpf, M.~Eibl, W.~Wieser, T.~Klein, and R.~Huber, \enquote{A time-encoded
  technique for fibre-based hyperspectral broadband stimulated {R}aman
  microscopy,} {\protect\JournalTitle{Nature Communications}} \textbf{6}, 6784
  (2015).

\bibitem{eigenwillig2013picosecond}
C.~M. Eigenwillig, W.~Wieser, S.~Todor, B.~R. Biedermann, T.~Klein,
  C.~Jirauschek, and R.~Huber, \enquote{Picosecond pulses from wavelength-swept
  continuous-wave {F}ourier domain mode-locked lasers,}
  {\protect\JournalTitle{Nature Communications}} \textbf{4}, 1848 (2013).

\bibitem{mao2009HighOutputPower}
Y.~Mao, C.~Flueraru, S.~Sherif, and S.~Chang, \enquote{High performance
  wavelength-swept laser with mode-locking technique for optical coherence
  tomography,} {\protect\JournalTitle{Optics Communications}} \textbf{282},
  88--92 (2009).

\bibitem{qiao2017VCSELReview}
P.~Qiao, K.~T. Cook, K.~Li, and C.~J. Chang-Hasnain, \enquote{Wavelength-swept
  {VCSEL}s,} {\protect\JournalTitle{IEEE Journal of Selected Topics in Quantum
  Electronics}} \textbf{23}, 1--16 (2017).

\bibitem{potsaid2012mems}
B.~Potsaid, V.~Jayaraman, J.~G. Fujimoto, J.~Jiang, P.~J. Heim, and A.~E.
  Cable, \enquote{{MEMS} tunable {VCSEL} light source for ultrahigh speed
  60k{H}z-1{MH}z axial scan rate and long range centimeter class {OCT}
  imaging,} in \emph{Optical Coherence Tomography and Coherence Domain Optical
  Methods in Biomedicine XVI,}  vol. 8213 (International Society for Optics and
  Photonics, 2012), p. 82130M.

\bibitem{bonesi2014akinetic}
M.~Bonesi, M.~Minneman, J.~Ensher, B.~Zabihian, H.~Sattmann, P.~Boschert,
  E.~Hoover, R.~Leitgeb, M.~Crawford, and W.~Drexler, \enquote{Akinetic
  all-semiconductor programmable swept-source at 1550 nm and 1310 nm with
  centimeters coherence length,} {\protect\JournalTitle{Optics Express}}
  \textbf{22}, 2632--2655 (2014).

\bibitem{pfeiffer2017analysis}
T.~Pfeiffer, W.~Draxinger, W.~Wieser, T.~Klein, M.~Petermann, and R.~Huber,
  \enquote{Analysis of {FDML} lasers with meter range coherence,} in
  \emph{Optical Coherence Tomography and Coherence Domain Optical Methods in
  Biomedicine XXI,}  vol. 10053 (International Society for Optics and
  Photonics, 2017), p. 100531T.

\bibitem{kraetschmer2009ultrastable}
T.~Kraetschmer and S.~T. Sanders, \enquote{Ultrastable {F}ourier domain mode
  locking observed in a laser sweeping 1363.8 -- 1367.3 nm,} in
  \emph{Conference on Lasers and Electro-Optics,}  (Optical Society of America,
  2009), p. CFB4.

\bibitem{pfeiffer2018ultra}
T.~Pfeiffer, M.~Petermann, W.~Draxinger, C.~Jirauschek, and R.~Huber,
  \enquote{Ultra low noise {F}ourier domain mode locked laser for high quality
  megahertz optical coherence tomography,} {\protect\JournalTitle{Biomedical
  Optics Express}} \textbf{9}, 4130--4148 (2018).

\bibitem{hartl2001ultrahigh}
I.~Hartl, X.~D. Li, C.~Chudoba, R.~K. Ghanta, T.~H. Ko, J.~G. Fujimoto, J.~K.
  Ranka, and R.~S. Windeler, \enquote{Ultrahigh-resolution optical coherence
  tomography using continuum generation in an air--silica microstructure
  optical fiber,} {\protect\JournalTitle{Optics Letters}} \textbf{26}, 608--610
  (2001).

\bibitem{nishizawa2004real}
N.~Nishizawa, Y.~Chen, P.~Hsiung, E.~P. Ippen, and J.~G. Fujimoto,
  \enquote{Real-time, ultrahigh-resolution, optical coherence tomography with
  an all-fiber, femtosecond fiber laser continuum at 1.5 $\mu$m,}
  {\protect\JournalTitle{Optics Letters}} \textbf{29}, 2846--2848 (2004).

\bibitem{huber2006buffered}
R.~Huber, D.~C. Adler, and J.~G. Fujimoto, \enquote{Buffered {F}ourier domain
  mode locking: unidirectional swept laser sources for optical coherence
  tomography imaging at 370,000 lines/s,} {\protect\JournalTitle{Optics
  Letters}} \textbf{31}, 2975--2977 (2006).

\bibitem{tsai2009frequency}
T.-H. Tsai, C.~Zhou, D.~C. Adler, and J.~G. Fujimoto, \enquote{Frequency comb
  swept lasers,} {\protect\JournalTitle{Optics Express}} \textbf{17},
  21257--21270 (2009).

\bibitem{wan2017time}
M.~Wan, F.~Li, X.~Feng, X.~Wang, Y.~Cao, B.-O. Guan, D.~Huang, J.~Yuan, and
  P.~K.~A. Wai, \enquote{Time and {F}ourier domain jointly mode locked
  frequency comb swept fiber laser,} {\protect\JournalTitle{Optics Express}}
  \textbf{25}, 32705--32712 (2017).

\bibitem{lippok2019extended}
N.~Lippok, M.~Siddiqui, B.~J. Vakoc, and B.~E. Bouma, \enquote{Extended
  coherence length and depth ranging using a {F}ourier-domain mode-locked
  frequency comb and circular interferometric ranging,}
  {\protect\JournalTitle{Physical Review Applied}} \textbf{11}, 014018 (2019).

\bibitem{pimenov2017dispersive}
A.~Pimenov, S.~Slepneva, G.~Huyet, and A.~G. Vladimirov, \enquote{Dispersive
  time-delay dynamical systems,} {\protect\JournalTitle{Physical Review
  Letters}} \textbf{118}, 193901 (2017).

\bibitem{slepneva2013dynamics}
S.~Slepneva, B.~Kelleher, B.~O'Shaughnessy, S.~P. Hegarty, A.~G. Vladimirov,
  and G.~Huyet, \enquote{Dynamics of {F}ourier domain mode-locked lasers,}
  {\protect\JournalTitle{Optics Express}} \textbf{21}, 19240--19251 (2013).

\bibitem{slepneva2019convective}
S.~Slepneva, B.~O'Shaughnessy, A.~G. Vladimirov, S.~Rica, E.~A. Viktorov, and
  G.~Huyet, \enquote{Convective {N}ozaki-{B}ekki holes in a long cavity {OCT}
  laser,} {\protect\JournalTitle{Optics Express}} \textbf{27}, 16395--16404
  (2019).

\bibitem{li2017eckhaus}
F.~Li, K.~Nakkeeran, J.~N. Kutz, J.~Yuan, Z.~Kang, X.~Zhang, and P.~K.~A. Wai,
  \enquote{Eckhaus instability in the {F}ourier-domain mode locked fiber laser
  cavity,} {\protect\JournalTitle{arXiv preprint arXiv:1707.08304}}  (2017).

\bibitem{jirauschek2009theoretical}
C.~Jirauschek, B.~Biedermann, and R.~Huber, \enquote{A theoretical description
  of {F}ourier domain mode locked lasers,} {\protect\JournalTitle{Optics
  Express}} \textbf{17}, 24013--24019 (2009).

\bibitem{jirauschek2017efficient}
C.~Jirauschek and R.~Huber, \enquote{Efficient simulation of the swept-waveform
  polarization dynamics in fiber spools and {F}ourier domain mode-locked
  ({FDML}) lasers,} {\protect\JournalTitle{Journal of the Optical Society of
  America B}} \textbf{34}, 1135--1146 (2017).

\bibitem{agrawal1989self}
G.~P. Agrawal and N.~A. Olsson, \enquote{Self-phase modulation and spectral
  broadening of optical pulses in semiconductor laser amplifiers,}
  {\protect\JournalTitle{IEEE Journal of Quantum Electronics}} \textbf{25},
  2297--2306 (1989).

\bibitem{bekki1985formations}
N.~Bekki and K.~Nozaki, \enquote{Formations of spatial patterns and holes in
  the generalized {G}inzburg-{L}andau equation,} {\protect\JournalTitle{Physics
  Letters A}} \textbf{110}, 133--135 (1985).

\bibitem{chate1992stability}
H.~Chat{\'e} and P.~Manneville, \enquote{Stability of the {B}ekki-{N}ozaki hole
  solutions to the one-dimensional complex {G}inzburg-{L}andau equation,}
  {\protect\JournalTitle{Physics Letters A}} \textbf{171}, 183--188 (1992).

\bibitem{chate1994spatiotemporal}
H.~Chat{\'e}, \enquote{Spatiotemporal intermittency regimes of the
  one-dimensional complex {G}inzburg-{L}andau equation,}
  {\protect\JournalTitle{Nonlinearity}} \textbf{7}, 185--204 (1994).

\bibitem{kolb2016megahertz}
J.~P. Kolb, T.~Klein, M.~Eibl, T.~Pfeiffer, W.~Wieser, and R.~Huber,
  \enquote{Megahertz {FDML} laser with up to 143nm sweep range for ultrahigh
  resolution {OCT} at 1050nm,} in \emph{Optical Coherence Tomography and
  Coherence Domain Optical Methods in Biomedicine XX,}  vol. 9697
  (International Society for Optics and Photonics, 2016), p. 969703.

\bibitem{cassioli2000time}
D.~Cassioli, S.~Scotti, and A.~Mecozzi, \enquote{A time-domain computer
  simulator of the nonlinear response of semiconductor optical amplifiers,}
  {\protect\JournalTitle{IEEE Journal of Quantum Electronics}} \textbf{36},
  1072--1080 (2000).

\bibitem{pfeiffer2016flexible}
T.~Pfeiffer, W.~Wieser, T.~Klein, M.~Petermann, J.-P. Kolb, M.~Eibl, and
  R.~Huber, \enquote{Flexible {A}-scan rate{ MHz OCT}: {C}omputational
  downscaling by coherent averaging,} in \emph{Optical Coherence Tomography and
  Coherence Domain Optical Methods in Biomedicine XX,}  vol. 9697
  (International Society for Optics and Photonics, 2016), p. 96970S.

\bibitem{kano1993reduction}
F.~Kano, T.~Yamanaka, N.~Yamamoto, Y.~Yoshikuni, H.~Mawatari, Y.~Tohmori,
  M.~Yamamoto, and K.~Yokoyama, \enquote{Reduction of linewidth enhancement
  factor in {InGaAsP-InP} modulation-doped strained multiple-quantum-well
  lasers,} {\protect\JournalTitle{IEEE Journal of Quantum Electronics}}
  \textbf{29}, 1553--1559 (1993).

\bibitem{zah1994high}
C.-E. Zah, R.~Bhat, B.~N. Pathak, F.~Favire, W.~Lin, M.~C. Wang, N.~C.
  Andreadakis, D.~M. Hwang, M.~A. Koza, T.-P. Lee, Z.~Wang, D.~Darby,
  D.~Flanders, and J.~J. Hsieh, \enquote{{High-performance uncooled
  $1.3$-$\mathrm{\mu}$m $\mathrm{Al_xGa_yIn_{1-x-y}As/InP}$ strained-layer
  quantum-well lasers for subscriber loop applications},}
  {\protect\JournalTitle{IEEE Journal of Quantum Electronics}} \textbf{30},
  511--523 (1994).

\bibitem{tiemeijer1991dependence}
L.~F. Tiemeijer, P.~J.~A. Thijs, P.~J. de~Waard, J.~J.~M. Binsma, and T.~v.
  Dongen, \enquote{{Dependence of polarization, gain, linewidth enhancement
  factor, and K factor on the sign of the strain of InGaAs/InP strained-layer
  multiquantum well lasers},} {\protect\JournalTitle{Applied Physics Letters}}
  \textbf{58}, 2738--2740 (1991).

\bibitem{thiis1994progress}
P.~J.~A. Thiis, L.~F. Tiemeijer, J.~J.~M. Binsma, and T.~van Dongen,
  \enquote{Progress in long-wavelength strained-layer {InGaAs(P)} quantum-well
  semiconductor lasers and amplifiers,} {\protect\JournalTitle{IEEE Journal of
  Quantum Electronics}} \textbf{30}, 477--499 (1994).

\bibitem{dorren2004all}
H.~J.~S. Dorren, X.~Yang, A.~K. Mishra, Z.~Li, H.~Ju, H.~de~Waardt, G.-D. Khoe,
  T.~Simoyama, H.~Ishikawa, H.~Kawashima, and T.~Hasama, \enquote{All-optical
  logic based on ultrafast gain and index dynamics in a semiconductor optical
  amplifier,} {\protect\JournalTitle{IEEE Journal of Selected Topics in Quantum
  Electronics}} \textbf{10}, 1079--1092 (2004).

\bibitem{huang2011optimized}
X.~Huang, Z.~Zhang, C.~Qin, Y.~Yu, and X.~Zhang, \enquote{Optimized
  quantum--well semiconductor optical amplifier for {RZ-DPSK} signal
  regeneration,} {\protect\JournalTitle{IEEE Journal of Quantum Electronics}}
  \textbf{47}, 819--826 (2011).

\bibitem{grigoryan2006soa}
V.~S. Grigoryan, M.~Shin, P.~Devgan, J.~Lasri, and P.~Kumar,
  \enquote{{SOA}-based regenerative amplification of phase-noise-degraded
  {DPSK} signals:{ D}ynamic analysis and demonstration,}
  {\protect\JournalTitle{Journal of Lightwave Technology}} \textbf{24},
  135--142 (2006).

\bibitem{zilkie2008time}
A.~J. Zilkie, J.~Meier, M.~Mojahedi, A.~S. Helmy, P.~J. Poole, P.~Barrios,
  D.~Poitras, T.~J. Rotter, C.~Yang, A.~Stintz, K.~J. Malloy, P.~W.~E. Smith,
  and J.~S. Aitchison, \enquote{Time-resolved linewidth enhancement factors in
  quantum dot and higher-dimensional semiconductor amplifiers operating at 1.55
  $\mu$m,} {\protect\JournalTitle{Journal of Lightwave Technology}}
  \textbf{26}, 1498--1509 (2008).

\bibitem{mecozzi1997saturationFWM}
A.~Mecozzi and J.~M{\o}rk, \enquote{Saturation effects in nondegenerate
  four-wave mixing between short optical pulses in semiconductor laser
  amplifiers,} {\protect\JournalTitle{IEEE Journal of Selected Topics in
  Quantum Electronics}} \textbf{3}, 1190--1207 (1997).

\bibitem{mecozzi1997saturation}
A.~Mecozzi and J.~M{\o}rk, \enquote{Saturation induced by picosecond pulses in
  semiconductor optical amplifiers,} {\protect\JournalTitle{Journal of the
  Optical Society of America B}} \textbf{14}, 761--770 (1997).

\bibitem{storkfelt1991measurement}
N.~Storkfelt, B.~Mikkelsen, D.~S. Olesen, M.~Yamaguchi, and K.~E. Stubkjaer,
  \enquote{Measurement of carrier lifetime and linewidth enhancement factor for
  1.5-$\mu$m ridge-waveguide laser amplifier,} {\protect\JournalTitle{IEEE
  Photonics Technology Letters}} \textbf{3}, 632--634 (1991).

\bibitem{clavero2005all}
R.~Clavero, F.~Ramos, J.~M. Martinez, and J.~Marti, \enquote{All-optical
  flip-flop based on a single {SOA-MZI},} {\protect\JournalTitle{IEEE Photonics
  Technology Letters}} \textbf{17}, 843--845 (2005).

\bibitem{tiemeijer1996self}
L.~F. Tiemeijer, P.~J.~A. Thijs, T.~van Dongen, J.~J.~M. Binsma, and E.~J.
  Jansen, \enquote{Self-phase modulation coefficient of multiple-quantum-well
  optical amplifiers,} {\protect\JournalTitle{IEEE Photonics Technology
  Letters}} \textbf{8}, 876--878 (1996).

\bibitem{bilenca2006numerical}
A.~Bilenca, S.~H. Yun, G.~J. Tearney, and B.~E. Bouma, \enquote{Numerical study
  of wavelength-swept semiconductor ring lasers: the role of refractive-index
  nonlinearities in semiconductor optical amplifiers and implications for
  biomedical imaging applications,} {\protect\JournalTitle{Optics Letters}}
  \textbf{31}, 760--762 (2006).

\bibitem{girard2011soa}
S.~L. Girard, M.~Pich{\'e}, H.~Chen, G.~W. Schinn, W.-Y. Oh, and B.~E. Bouma,
  \enquote{{SOA} fiber ring lasers: {S}ingle-versus multiple-mode oscillation,}
  {\protect\JournalTitle{IEEE Journal of Selected Topics in Quantum
  Electronics}} \textbf{17}, 1513--1520 (2011).

\bibitem{wang2007temporal}
J.~Wang, A.~Maitra, C.~G. Poulton, W.~Freude, and J.~Leuthold,
  \enquote{Temporal dynamics of the alpha factor in semiconductor optical
  amplifiers,} {\protect\JournalTitle{Journal of Lightwave Technology}}
  \textbf{25}, 891--900 (2007).

\bibitem{mork1997theory}
J.~M{\o}rk and A.~Mecozzi, \enquote{Theory of nondegenerate four-wave mixing
  between pulses in a semiconductor waveguide,} {\protect\JournalTitle{IEEE
  Journal of Quantum Electronics}} \textbf{33}, 545--555 (1997).

\bibitem{SMF28}
{Corning}, \enquote{{Corning SMF-28 optical fiber product information},}
  \url{http://www.princetel.com/datasheets/smf28e.pdf}.

\bibitem{wieser2012chromatic}
W.~Wieser, G.~Palte, C.~M. Eigenwillig, B.~R. Biedermann, T.~Pfeiffer, and
  R.~Huber, \enquote{Chromatic polarization effects of swept waveforms in
  {FDML} lasers and fiber spools,} {\protect\JournalTitle{Optics Express}}
  \textbf{20}, 9819--9832 (2012).

\bibitem{HI1060}
{Corning}, \enquote{Corning {HI} 1060 \& {RC HI} 1060 specialty optical
  fibers,}
  \url{https://www.corning.com/media/worldwide/csm/documents/HI\%201060\%20Specialty\%20Fiber\%20PDF.pdf}.

\bibitem{todor2011instantaneous}
S.~Todor, B.~Biedermann, W.~Wieser, R.~Huber, and C.~Jirauschek,
  \enquote{Instantaneous lineshape analysis of {F}ourier domain mode-locked
  lasers,} {\protect\JournalTitle{Optics Express}} \textbf{19}, 8802--8807
  (2011).

\bibitem{todor2012balance}
S.~Todor, B.~Biedermann, R.~Huber, and C.~Jirauschek, \enquote{Balance of
  physical effects causing stationary operation of {F}ourier domain mode-locked
  lasers,} {\protect\JournalTitle{Journal of the Optical Society of America B}}
  \textbf{29}, 656--664 (2012).

\end{thebibliography}

\end{document}